\documentclass[10pt,a4paper]{article} 
\usepackage{lipsum}
\usepackage{url}
\usepackage[nochapters,subfig]{classicthesis} 

\usepackage{amsthm}
\usepackage{amsfonts}
\usepackage{amsmath}
\usepackage{amssymb}
\usepackage{nicefrac}
\usepackage{graphicx}
\usepackage{booktabs}
\usepackage{footnote}
\usepackage[english]{babel}
\usepackage[T1]{fontenc} 
\usepackage{microtype}
\usepackage{afterpage}
\usepackage[smaller]{acronym}
\usepackage[affil-it]{authblk}
\usepackage{subfig}
\usepackage{xcolor}

\newcounter{thmctr}
\newenvironment{lmm}{
   \par\medskip\noindent
   \refstepcounter{thmctr}
   \spacedlowsmallcaps{Lemma \thethmctr}
   \begin{it}
   }
   {\end{it}
   \par\smallskip}  

\newenvironment{prp}{
   \par\medskip\noindent
   \refstepcounter{thmctr}
   \spacedlowsmallcaps{Proposition \thethmctr}
   \begin{it}
   }
   {\end{it}
   \par\smallskip}  

\theoremstyle{remark}
\newtheorem{rmk}{Remark}
\newtheorem{dfn}{Definition}
\newcommand{\ie}{{i.\,e.\ }}
\newcommand{\eg}{{e.\,g.\ }}
\newcommand{\myw}{\ensuremath{\omega}}

\newcommand{\srlJ}{\ensuremath{J^{\text{\tiny SRL}}}}
\newcommand{\srlPi}{\ensuremath{\pi^{\text{\tiny SRL}}}}

\newcommand{\parJ}{\ensuremath{J}}
\newcommand{\parPi}{\ensuremath{\pi}}
\newcommand{\blkP}{\ensuremath{P_\varepsilon}}

\newcommand{\infP}{\ensuremath{P_{\varepsilon,\infty}}}
\newcommand{\infJ}{\ensuremath{J_{\varepsilon,\infty}}}
\newcommand{\infPi}{\ensuremath{\pi_{\varepsilon,\infty}}}
\usepackage{dsfont}
\newcommand{\ind}[1]{\mathds{1}_{\{#1\}}}

\begin{document}

    \title{\rmfamily\normalfont\spacedallcaps{On the blockage problem
        and the non-analyticity of the current for the parallel TASEP
        on a ring}} 
    \author{Benedetto Scoppola}
    \author{Carlo Lancia}
    \author{Riccardo Mariani}
    \affil{Universit\`a di Roma `Tor Vergata'}

    \date{\today}

    \maketitle

    \acrodef{TASEP}{Totally Asymmetric Simple Exclusion Process}
    \acrodef{PCA}{Probabilistic Cellular Automata}

    \begin{abstract}
      \noindent The \acl{TASEP}~(\acs{TASEP})
      is an important example of a particle system driven by an irreversible Markov chain.
      In this paper we give a simple yet rigorous derivation of the chain stationary measure
      in the case of parallel updating rule. In this parallel framework we then consider the
      \emph{blockage problem} (aka slow bond problem).
      We find the exact expression of the current 
      for an arbitrary blockage intensity $\varepsilon$ in the case of
      the so-called \emph{rule-184 cellular automaton}, i.e. a parallel \acs{TASEP} where at each step
      all particles free-to-move are actually moved.
      Finally, we investigate through numerical experiments the conjecture that for parallel updates other than
      rule-184 the current may be non-analytic in the blockage intensity around the value $\varepsilon = 0$.
      
      \smallskip

      \noindent\textsc{keywords:} Parallel \textsc{tasep}; Blockage Problem; Current
      
      \smallskip

      \noindent\textsc{msc2010:} 60J10, 37B15, 60K30
    \end{abstract}
       

    \section{Introduction}
    \label{sec:intro}
    
    The \ac{TASEP} is one of the more popular
    example of discrete particle system driven by a Markov irreversible
    dynamics~\cite{liggett1985interacting,liggett1999stochastic}. 
    In finite space\footnote{Although it is possible
      to define the \ac{TASEP} on the whole $\mathbb{Z}$ by considering continuous time,
      in this paper we will consider only finite space (and discrete time).}
    the system can be defined either on a discrete
    segment $\Lambda = \{1,2, \ldots ,2L\}$ or on a discrete circle by
    imposing periodic boundary condition to $\Lambda$.
    A configuration $\sigma \in \{0,1\}^\Lambda$
    can be viewed as a set of particles \emph{living} in $\Lambda$.
    According to this map, $\sigma_i = 1$ means that the $i$th site is occupied by a particle,
    whereas if $\sigma_i = 0$ then the $i$th site is a \emph{hole}, \ie it is an empty site.
    
    \ac{TASEP} may be formulated either as a serial
    or a parallel dynamics.
    The serial \ac{TASEP} chooses an occupied site $i$
    uniformly at random;
    if the right-neighbouring site, \ie site\footnote{Please bear in
      mind that the right-neighbour is actually site $(i+1)\!\! \mod n$ in view of the
      periodic boundary conditions.} $i+1$, is a hole then the
    selected particle jumps to the $(i+1)$th site;
    conversely the $(i+1)$th site is occupied by another particle,
    then the configuration does not change and a new iteration starts
    over.
    The parallel \ac{TASEP} selects instead \emph{all} the particles having an unoccupied
    right-neighbouring site, 
    but those actually advancing to the empty site are chosen
    according to a binomial rule, \ie each particle actually advances with an independent probability $p$.
    Both serial and parallel dynamics are clearly irreversible.
    
    Despite its simplicity, the serial \ac{TASEP} exhibits many interesting features.
    For example, the stationary measure is uniform on a discrete circle since the
    transition matrix is doubly Markov.
    On the finite segment the stationary state depends on the probability of a particle
    entering or exiting the system through the left and 
    right extremes of the segment, respectively (see
    \cite{derrida1993exact,schutz1993phase}).
    In this case the system exhibits a non-trivial behaviour.
    
    The study of \ac{TASEP} typically focuses on the \emph{current}, defined as the probability that, 
    under stationary conditions, the $i$th site is a particle and
    the $(i+1)$th is a hole.
    This quantity is important because it measures the tendency of the system to exhibit congestion,
    \ie the tendency to form long sequences of clustered particles that are not free to move.
    Remarkably, the current does not depend on $i$ and can be exactly computed for both the serial \ac{TASEP}
    defined on the segment and on the discrete circle.
    In particular, on the discrete circle the current depends only on the number
    of particles in the system and it is maximum (and equal to $\nicefrac{1}{4}$ 
    in the limit of $L\to\infty$) when the system is half-filled, \ie
    when $L$ particles lie on the circle $\Lambda = \{1,2,...,2L\}$.
    
    Parallel \ac{TASEP} was first studied on a ring in~\cite{nagel1992cellular} as a
    collective dynamics for freeway traffic. The model was extended
    in~\cite{schreckenberg1995discrete} to the case of particles
    performing jumps of length corresponding to their velocity. In the
    case of jumps limited to next-neighbouring sites (velocity 1), the authors derived
    through a mean-field approximation the steady-state distribution of
    the model. However, their derivation is quite involved, requiring to
    prove that the first order mean-field approximation remains valid at all higher
    orders. 
    Parallel \ac{TASEP} on a ring with inhomogeneous jump-rates is studied
    in~\cite{evans1997exact}, where the stationary distribution of the
    model is given in a form equivalent
    to~\cite{schreckenberg1995discrete}.
    Ref.~\cite{schadschneider2002traffic} gives a review of parallel
    \ac{TASEP} on a ring with a focus on traffic applications.
    More recently, the steady-state of a parallel \ac{TASEP} on a ring
    where particles can jump one or two sites ahead has been characterised
    as a scalar-pair factorised and a matrix-product state~\cite{woelki2009exact}.
    
    Exact solution to parallel \ac{TASEP} with open boundary conditions is
    given both in~\cite{de1999exact} and~\cite{evans1999exact} in terms of
    a matrix product. An alternative, combinatorial 
    expression in term of Catalan numbers can be found in~\cite{duchi2008combinatorial}.
    Parallel \ac{TASEP} has been investigated also via Bethe Ansatz
    imposing both open and periodic boundary. The results obtained so far
    comprehend the expression of the evolute measure of the model at every
    time~\cite{povolotsky2006determinant,povolotsky2007determinant,mallick2011some}.
    
    All formulations of the stationary distribution of the parallel
    \ac{TASEP} (both with open and periodic boundary conditions) are considered quite involved
    in~\cite{woelki2013parallel}, where the author studies parallel
    \ac{TASEP} with open boundaries in terms of Motzkin paths.
    In this respect, our work offers a significant contribution, for it
    frames the steady state of parallel \ac{TASEP} in a simple yet fruitful formulation.
    
    A very natural question in the study of the irreversible system is 
    whether the effect of small perturbations in the dynamics has local effects, 
    as in the case of reversible system far from critical points, or
    global effects. In the case of the \ac{TASEP} this question has been
    investigated by imposing the so-called {\it blockage}
    (see~\cite{janowsky1992finite,janowsky1994exact}), that is to say, 
    in a defined point
    (say, without loss of 
    generality, in the point $2L$ of the circle)
    the probability to jump to the empty
    site $1$ is reduced by a factor $1-\varepsilon$ with $\varepsilon>0$.
    In terms of freeway traffic, blockage is a synonym for bottleneck and
    its effects are typically studied by a simulative approach; one of the
    first studies of this kind is about the so-called rule-184 cellular
    automaton~\cite{yukawa1994dynamical}, which is nothing else than a
    parallel \ac{TASEP} with deterministic updates, \ie a parallel \ac{TASEP}
    with probability $p=1$.
    
    Answering the question about global
    effects on the system here simply reduces to the evaluation of the current;
    if a blockage in a single point influences the value of the current then
    its effects are obviously global.
    For a long time it has been unclear whether the presence of a
    blockage of intensity $\varepsilon$
    had global effects for all $\varepsilon>0$.
    It is conjectured in~\cite{costin2012blockage}
    that the current decreases, for
    small $\varepsilon$, with a non-analytic dependence on $\varepsilon$.
    Numerical evaluation
    of the current may suggest also the existence of a critical value
    $\varepsilon_c>0$ such that the current does
    not change for $\varepsilon<\varepsilon_c$.
    Only very recently it has been proved~\cite{basu2014last}
    that for $\Lambda = \mathbb{Z}$ and continuous time, it is $\varepsilon_c=0$.
    However, the conjecture about the non-analyticity of the current around
    $\varepsilon=0$ still remains unproved.
    
    In this paper we study the parallel \ac{TASEP} dynamics,
    where at each step each particle
    followed by an empty site has a finite probability $p$ to jump.
    We show that this model has similar features with
    respect to the serial
    \ac{TASEP}; in particular, considering the blockage
    problem, we see numerically that for $p<1$ the behaviour of the current
    is very similar to the standard \ac{TASEP} case, because for
    small $\varepsilon$ the
    current suggests a non-analytic dependence on $\varepsilon$
    around $\varepsilon=0$.
    On the other hand, for $p=1$, we prove that the system with blockage
    is exactly solvable, and that
    the current is analytic as a function of $0 \le \varepsilon \le 1$.
    Regarding the existence of a critical value $\varepsilon_c$,
    it is likely that the very same argument
    of~\cite{basu2014last} may be extended to the case of the discrete
    circle for both serial and parallel dynamics.
    We hence expect that, except for the case $p=1$, where the
    current as a function of $\varepsilon$ has a finite first derivative,
    the current is non-analytic around $\varepsilon=0$ for all $p<1$.
    
    The paper is organised as follows: in Section~\ref{Intro} we 
    define our parallel irreversible dynamics
    on the circle, and we compute its stationary distribution. The computation of the
    stationary measure is quite simple, being based on the idea of
    \emph{dynamical reversibility}. 
    In Section~\ref{current}
    we explicitly compute the current of parallel \ac{TASEP}
    making use of the exact expression of the stationary measure, and we
    find with relative ease the expression already known
    in the literature.
    In Section~\ref{block} we investigate the blockage problem for the parallel \ac{TASEP},
    and we solve it in the particular case of $p=1$, that is to say,
    the 184-rule with a blockage of intensity $\varepsilon$
    in a single site. The stochasticity of the resulting process sits in this case
    only in the blockage, since in all the other sites the dynamics is deterministic.
    Finally, in Section~\ref{num}
    we present some numerical simulations and discuss a conjecture inspired by~\cite{costin2012blockage}
    on the non-analyticity of the current
    around $\varepsilon = 0$.
    
    \section{The parallel TASEP}
    \label{Intro}
    
    In this paper we study the parallel \ac{TASEP} on a discrete circle, 
    i.e.\ on the set $\Lambda=\{1,2,...,2L\}$
    with periodic boundary conditions. 
    A configuration $\sigma$ is any element of the set
    $\{0,1\}^\Lambda$, which can be though of as a $|\Lambda|$-dimensional
    vector; the $i$th component of the vector is $\sigma_i$.
    We say that $i$th site is a \emph{particle} whenever $\sigma_i=1$,
    while the site is a \emph{hole} if $\sigma_i=0$.
    We allow for \emph{negative indexing} of the configuration $\sigma$ under the
    agreement that $\sigma_{-i} = \sigma_{2L-i+1}$
    
    A particle in site $i$ is said to be \emph{free to move} whenever the
    site $i+1$ is a hole.
    Particles move clockwise by exchanging position with the neighbouring
    hole. In other words, 
    moving a particle means replacing the couple
    $\sigma_i=1, \sigma_{i+1}=0$ with $\sigma_i=0,\sigma_{i+1}=1$.
    As a consequence, holes move counterclockwise.
    The symbols $\bullet$ and $\times$ are used as convenient shorthands
    for particles and holes, respectively. Using this notation, the process of moving a
    particle can be sketched as $\bullet \times \to
    \times \bullet$.
    
    Each configuration $\sigma$ can be decomposed into \emph{particle trains}, \ie maximal sequences of
    particles lying in adjacent sites. The \emph{engine} of a
    train is the element of the sequence with highest
    index possible; conversely, the \emph{caboose} is the element with lowest
    index\footnote{Except if the train has length bigger than~1
      and one of the particles composing it
      occupies site~1. In this case engine and caboose are the elements with
      smallest and highest index, respectively.}.
    Engines are the only particles that can move across a single
    iteration of the dynamics. Whenever an engine is moved, it may either form a
    new train of length 1 ($\bullet \bullet \times \times \to
    \bullet \times \bullet \times$) or become the caboose of another train 
    ($\bullet \bullet \times \bullet \to
    \bullet \times \bullet \bullet$).
    
    Given a configuration $\sigma$, let $m(\sigma)=\sum_{i=1}^{2L}\sigma_i$
    be the number of particles living in $\Lambda$. In what follows we
    assume that $m(\sigma)\le L$. 
    We indicate by $l(\sigma) \le m(\sigma)$ the number of engines (or
    trains) in the configuration $\sigma$.
    
    A transition from the configuration $\sigma$ to the configuration $\tau$
    is weighed according to the following rule:
    \begin{equation}\label{pesi}
      w(\sigma,\tau)=
      \begin{cases}
        \myw^n&\text{if } \tau \text{ can be reached by moving } n \text{ particles in } \sigma,\\
        0 & \text{otherwise},
      \end{cases}
    \end{equation}
    where $\myw>0$ is a positive parameter measuring the tendency of a
    free particle to move.
    The resulting transition probabilities are
    \begin{equation}\label{4.a}
      P(\sigma,\tau)=\frac{w(\sigma,\tau)}{w(\sigma)} \,,
    \end{equation}
    where 
    \begin{equation}\label{4}
      w(\sigma)=\sum_\tau w(\sigma,\tau)=\sum_{k=0}^{l(\sigma)} {l(\sigma)\choose k} \myw^k=
      (1+\myw)^{l(\sigma)}.
    \end{equation}
    The dynamics clearly conserves the number of particles in $\Lambda$,
    \ie $m(\sigma) = m(\tau) = m$. 
    
    For small values of \myw{} (\eg $\myw = O(\nicefrac{1}{L})$), the
    parallel \ac{TASEP} is similar to a serial dynamics. For finite values
    of \myw{}, the dynamics is truly parallel, in the sense that each
    free-to-move particle advances to the empty neighbouring site with
    independent probability
    $p=\nicefrac{\myw}{1+\myw}$.
    In the limit of $\myw \to \infty$, $p=1$ and all the free particles simultaneously 
    move at each step (rule-184 cellular automata).
    
    The Markov chain~\eqref{4.a} is manifestly irreversible.
    However, it is still possible to compute the stationary distribution
    of the chain with relative ease as the chain satisfies the {\it global balance
      principle}~\cite{lancia2013equilibrium} 
    \begin{equation}\label{2}
      \sum_\tau w(\sigma,\tau)=\sum_{\tau^\prime} w(\tau^\prime,\sigma)\,.
    \end{equation}
    
    Equation~\eqref{2} is fulfilled simply because the final
    configurations $\tau$ at the left hand side of~\eqref{2} can be mapped
    one-to-one onto the initial configurations $\tau^\prime$ at the right hand
    side in such a way that $w(\sigma,\tau) = w(\tau^\prime, \sigma)$.
    More precisely, let $\tau$ be the configuration obtained from $\sigma$ by advancing
    the engine of some particle trains to
    the empty neighbouring sites; similarly, let $\tau^\prime$ the
    configuration obtained from $\sigma$ by detaching the caboose of the
    very same set of particle trains; clearly, $w(\sigma,\tau) = w(\tau^\prime, \sigma)$.
    
    Equation~\eqref{2} is known in the literature also by the name of {\it dynamical
      reversibility} and guarantees that the stationary measure of the
    chain is $\pi(\sigma)=\nicefrac{w(\sigma)}{W}$, where $W=\sum_\sigma w(\sigma)$.
    In fact,
    \begin{align}
      \label{3}
      \sum_\sigma\pi(\sigma)P(\sigma,\tau)=\sum_\sigma\frac{w(\sigma)}{W}\frac{w(\sigma,\tau)}{w(\sigma)}
      &=\sum_\sigma\frac{w(\sigma,\tau)}{W}\nonumber\\
      &=\sum_\sigma\frac{w(\tau,\sigma)}{W}=\frac{w(\tau)}{W}=\pi(\tau)\,.
    \end{align}

    Using~\eqref{4}, the stationary measure is
    \begin{equation}\label{4.1}
      \pi(\sigma)=
      \frac{(1+\myw)^{l(\sigma)}}{W}\,.
    \end{equation}
    
    \begin{rmk}
      When $\myw \to 0$, $\pi(\sigma)$ tends to the uniform value
      $\nicefrac{1}{ {2L \choose m(\sigma)} }$. In the limit of
      $\myw \to \infty$, $\pi(\sigma)$ tends instead to be uniform on all the
      configurations such that $l(\sigma)=m(\sigma)$ because in this
      regime $W=O(\myw^{m(\sigma)})$, so other
      configurations have a weight that is smaller by at least a factor
      $\nicefrac{1}{1+\myw}$.
    \end{rmk}

    \section{Current for the half-filled TASEP}
    \label{current}
    As mentioned in Section~\ref{sec:intro}, the value of the current
    $J$ in the serial \ac{TASEP} is the quantity 
    \begin{equation}
      \label{eq:Jserial}
      \srlJ = \lim_{\Lambda\to\infty}\srlPi(\sigma_i=1, \sigma_{i+1}=0)\,,
    \end{equation}
    where $\srlPi$ is the stationary distribution of the serial model and $i \in \Lambda$. 
    The event $\{\sigma_i=1, \sigma_{i+1}=0\}$  does not actually
    depend on the site $i$ but only on the number of particles $m$. 
    Therefore, an equivalent formulation for $\srlJ$ is the following:
    \begin{equation}
      \label{eq:Jserial2}
      \srlJ = \lim_{L \to \infty}
      \frac{\mathbb{E}_{\srlPi}[l(\sigma)]}{2L} \,,
    \end{equation}
    where $\mathbb{E}_{\srlPi}[\cdot]$ denotes expectation with respect to $\srlPi$.
    
    In what follows, we consider the \emph{half-filled}
    case $m = L$, i.e.\ the number of particles is exactly half the sites. 
    Then, for the serial \ac{TASEP},
    $$\srlJ = \srlPi(\sigma_i=1, \sigma_{i+1}=0) \simeq \srlPi(\sigma_i=1)\srlPi(\sigma_{i+1}=0) =
    \nicefrac{1}{4}$$
    since the stationary probability $\srlPi$ is uniform, and therefore the probability to 
    have the configuration $\sigma_i=1, \sigma_{i+1}=0$ tends, for large $\Lambda$, to the product 
    of the probabilities to have $\sigma_i=1$ and ${\sigma_{i+1}=0}$ (and they both equal 
    $\nicefrac{1}{2}$).
    
    In this section we focus on the current $\parJ$ in the parallel
    \ac{TASEP}. Analogously to~\eqref{eq:Jserial}-\eqref{eq:Jserial2}, let
    us define the current as
    \begin{equation}
      \label{eq:Jpar}
      \parJ = \parPi(\sigma_i=1, \sigma_{i+1}=0) = \lim_{L \to \infty} \frac{\mathbb{E}_{\parPi}[l(\sigma)]}{2L}\,,
    \end{equation}
    where $\mathbb{E}_{\parPi}[\cdot]$ denotes expectation with respect to $\parPi$.
    The following result holds:
    
    \begin{lmm}
      \label{lmm:parJ}
      For the parallel \ac{TASEP} \eqref{4.a}, the current $\parJ$ satisfies
      \begin{equation}\label{10.a}
        \parJ = \frac{1}{2}\ \frac{\sqrt{1 + \myw}}{1 + \sqrt{1 + \myw}}\,.
      \end{equation}
    \end{lmm}
    
    \begin{proof}
      Let us start by recasting $\parJ$ as
      \begin{equation}
        \label{6}
        \parJ = \lim_{L\to\infty}
        \frac{1}{2L}\ \frac{\sum_\sigma l(\sigma)(1 + \myw)^{l(\sigma)}}{\sum_\sigma(1 + \myw)^{l(\sigma)}} =
        \lim_{L\to\infty}\frac{1}{2L}\ \frac{\sum_{l=1}^{L} l\  
          n(l)(1 + \myw)^{l}}{\sum_{l=1}^{L} n(l)(1 + \myw)^{l}}\,,
      \end{equation}
      where $n(l)$ is the number of configurations $\sigma$ having $l(\sigma)=l$
      particles free to move. 
      Next, a formula for $n(l)$ is obtained as follows:
      first, we count the number of ways for dividing $L$ particles in
      $l$ distinct trains (and $L$ holes in $l$ trains);
      suppose that the first train of particles has length $l_1$, then we
      count all the ways for the site $i=1$ to fall within the first particle-train;
      at this stage a configuration is uniquely determined by the alternate sequence of particle and hole trains;
      it may also happen that the site $i=1$ falls within a hole train, and this is accounted for by multiplying everything by $2$.
      
      Since $L$ objects can be divided in $l$ ordered groups in ${L-1\choose l-1}$ ways, 
      \begin{equation}
        \label{7}
        n(l)=2\sum_{l_1=1}^{L-l+1}l_1{L-l_1-1\choose l-2}{L-1\choose l-1}\,.
      \end{equation}
      Substituting~\eqref{7} into~\eqref{6} yields
      \begin{equation}\label{8}
        \parJ = \lim_{L\to\infty}
        \frac{1}{2L}\  \frac{\sum_{l=1}^{L}\sum_{l_1=1}^{L-l+1}l_1{L-l_1-1\choose l-2}{L-1\choose l-1}
          \ l\  (1 + \myw)^{l}}{\sum_{l=1}^{L} \sum_{l_1=1}^{L-l+1}l_1{L-l_1-1\choose l-2}
          {L-1\choose l-1}(1 + \myw)^{l}}\,.
      \end{equation}
      In order to evaluate~\eqref{8}, let us write $l=\alpha L$, $l_1=\alpha_1 L$ and use the
      leading order approximation
      \begin{equation}\label{9.1}
        {n\choose \alpha n}\approx e^{nI(\alpha)}\,,
      \end{equation}
      where $I(\alpha)= -\alpha\ln\alpha-(1-\alpha)\ln(1-\alpha)$. Then,
      
      \begin{equation}\label{9}
        \parJ = \lim_{L\to\infty}
        \frac{1}{2L}\  L\frac{\int_0^1 d\alpha\int_0^{1-\alpha}d\alpha_1 \ \alpha \, \alpha_1 \, e^{L\left( (1-\alpha_1)I(\frac{\alpha}{1-\alpha_1})
              + I(\alpha) + \alpha\ln(1 + \myw) \right)}}{\int_0^1 d\alpha\int_0^{1-\alpha}d\alpha_1 \, \alpha_1 \, e^{L\left( (1 - \alpha_1)
              I(\frac{\alpha}{1-\alpha_1}) + I(\alpha)+\alpha\ln(1 + \myw)\right)}}\,.
      \end{equation}
      Let us define
      \[
      f(\alpha,\alpha_1)= (1-\alpha_1)I\left(\frac{\alpha}{1-\alpha_1}\right)+I(\alpha)+\alpha\ln(1+w)\,.
      \]
      Then, by the saddle-point method,
      \[\parJ = \lim_{L\to\infty}
      \left[\frac{1}{2}\bar\alpha+O\left(\frac{1}{L}\right)\right]\,,\] 
      where $\bar\alpha$ is the value
      of $\alpha$ that maximises $f(\alpha,\alpha_1)$.
      Since $f(\alpha,\alpha_1)$  is a decreasing function of $\alpha_1$,
      the choice $\alpha_1=0$ yields
      \begin{equation}
        \label{10}
        \bar\alpha = \frac{\sqrt{1 + \myw}}{1+\sqrt{1 + \myw}} \qquad\text{and}\qquad
        \parJ = \frac{1}{2}\ \frac{\sqrt{1 + \myw}}{1+\sqrt{1 + \myw}}\,.
      \end{equation}
    \end{proof}
    
    \begin{rmk}
      For $\myw=O (\nicefrac{1}{L} )$, the
      parallel \ac{TASEP} is a nearly serial dynamics and,
      as expected, $\parJ=\nicefrac{1}{4}$. 
      Moreover, $\parJ$ is an increasing function of $\myw$ and
      $\parJ=\nicefrac{1}{2}$ in the limit of $\myw\rightarrow\infty$.
    \end{rmk}

    \section{Blockage problem for the rule-184 cellular automata}
    \label{block}
    
    We have mentioned in Section~\ref{sec:intro} that a very interesting
    and difficult aspect in the study of serial \ac{TASEP} is the so-called
    blockage problem, which is defined as follows for the serial dynamics
    (see~\cite{costin2012blockage} and references therein). 
    At each step a particle is chosen uniformly at random and, if free to move, 
    it is moved to the next site with probability $1$ unless it occupies
    the site $2L$, in which case it is moved with probability
    $1-\varepsilon$.
    An explicit expression for the current is not yet known either for
    serial nor parallel \ac{TASEP}.
    
    Very recently, the current $\parJ$ in parallel \ac{TASEP} 
    was shown in~\cite{basu2014last} to be strictly less than the maximum
    value, i.e.\ $\nicefrac{1}{4}$, for each $\varepsilon > 0$.
    However, numerical evidences strongly suggest
    that the current stays very close to its
    maximum value up to some finite value of $\varepsilon$, see
    Section~\ref{num} below.
    Therefore, it may be that the current $\parJ$ as a function of
    $\varepsilon$ behaves as a very
    high-order polynomial around the value $\varepsilon=0$, 
    or that $J$ has an essential 
    singularity in $\varepsilon=0$ as conjectured in~\cite{costin2012blockage}.
    
    In this section we study the blockage problem through the parallel
    \ac{TASEP}~\eqref{pesi}-\eqref{4}
    modified in the following way:
    \begin{equation}
      \label{pesi2}
      w_\varepsilon(\sigma,\tau)=
      \begin{cases}
        \myw^n(1-\varepsilon\ind{\sigma_{2L}=1,\tau_{2L}=0}) &\text{if }
        \tau \text{ can be reached by moving }\\ &n \text{ particles in } \sigma,\\
        0 & \text{otherwise},
      \end{cases}
    \end{equation}
    where $\varepsilon \in (0,1)$ and $\ind{\cdot}$ is  the usual
    indicator function. Weights~\eqref{pesi2} are identical
    to~\eqref{pesi} save for they penalise by a factor $(1-\varepsilon)$
    the transitions $\bullet|\times \to \times|\bullet$ across the
    blockage, here represented by the symbol $| \,$.
    Let \ac{TASEP} with blockage be defined by the following transition probabilities:
    \begin{equation}\label{4.e}
      \blkP(\sigma,\tau)=\frac{w_\varepsilon(\sigma,\tau)}{w_\varepsilon(\sigma)}\,, 
    \end{equation}
    where
    \begin{equation}
      \label{eq:4.e.1}
      w_\varepsilon(\sigma) = \sum_\tau w_\varepsilon(\sigma, \tau) \,. 
    \end{equation}
    
    Due to the presence of the blockage, the global balance
    principle~\eqref{2} is no longer satisfied by the chain, and
    finding the stationary measure becomes extremely difficult.
    However, we will describe the current for finite values of $\myw$
    in Section~\ref{num} through numerical experiments.
    Apparently, the blockage problem for parallel \ac{TASEP}
    is similar to the serial case: the current seems to be almost constant 
    until $\varepsilon$ reaches
    a critical value, which is a decreasing function of $\myw$. 
    The possibility of rigorously proving 
    any result about this
    behaviour seems nevertheless to be as hard as for the serial dynamics.
    
    A happy exception is the regime $\myw\to\infty$, i.e.\ the rule-184
    cellular automata, where \emph{all} free particles move
    with probability $p=1$ except the particle at site $2L$, if any, which moves with
    probability $1-\varepsilon$. This case is easily solvable due to the
    circumstance that the dynamics preserves the so-called \emph{particle-hole
      symmetry}, defined next.
    
    \begin{dfn}
      A configuration $\sigma$ such that $\sigma_i = 1-\sigma_{-i}$ for each
      $i=1,2,\ldots,2L$ is said to be \emph{particle-hole symmetric}
      (abbrev.\ ph-symmetric). The
      set of particle-hole symmetric configurations is denoted by $PH$.
    \end{dfn}
    
    The next result states that in the limit of  $\myw\to\infty$ the parallel
    \ac{TASEP} with blockage preserves the particle-hole
    symmetry.
    
    \begin{lmm}
      \label{lmm:ph1}
      Consider the transition probabilities
      \begin{equation}
        \label{4.b}
        \infP(\sigma,\tau)=\lim_{\myw\to\infty} \blkP(\sigma,\tau)\,.
      \end{equation}
      For each $\varepsilon\in[0,1)$ and for each $\sigma\in PH$, 
      if $\infP(\sigma,\tau)>0$ then $\tau\in PH$.
    \end{lmm}
    
    \begin{proof}
      Let us fix a site $i\in\{2,3,\ldots,L\}$ and
      start considering the case $\sigma_i=1$. Due to the ph-symmetry, if
      $\sigma_i$ is an engine then also $\sigma_{-i-1}$ is an
      engine; since in the regime $\myw\to\infty$ the engines move with
      probability $1$, the sites $i$ and $-i-1$ will become holes, i.e.\
      $\tau_i=\tau_{-i-1}=0$, while the sites $i+1$ and $-i$ will be
      occupied by a particle, i.e.\ $\tau_{i+1}=\tau_{-i}=1$. Conversely,
      if $\sigma_i$ is not an engine then it must be $\sigma_{i+1}=1$,
      and so $\sigma_{-i} = \sigma_{-i-1} = 0$; therefore, site $i$
      will keep being occupied, while site $-i$ will continue being
      empty, that is to say, $\tau_i = 1$ and $\tau_{-i}=0$.
      The case $\sigma_i=0$ is completely analogous, but we have to
      distinguish between site $i$ being or not being the caboose of a
      hole train.
      We still have to check what happens across the blockage. The following
      four configurations are possible:
      \[\text{1.}\; \times\times | \bullet\bullet; \quad \text{2.}\; \times \bullet |
      \times\bullet; \quad \text{3.}\;\bullet \times | \bullet\times; \quad \text{4.}\;
      \bullet\bullet | \times\times. \]
      In the first case site $1$ remains occupied and site $2L$ empty,
      then $\tau_1=1$ and $\tau_{-1}=0$;
      in the third case site $1$ becomes empty and site $2L$ occupied,
      then $\tau_1=0$ and $\tau_{-1}=1$.
      Conversely, in the second and
      fourth case site $1$ becomes occupied and site $2L$ empty with
      probability $1-\varepsilon$, i.e.\ $\tau_1=1$ and $\tau_{-1}=0$, or with
      probability $\varepsilon$ the
      blockage acts on the particle at site $2L$, which remains occupied,
      i.e.\ $\tau_1=0$ and $\tau_{-1}=1$.
    \end{proof}
    
    The following two lemmas
    identify the set of recurrent states of the Markov chain $P_{\varepsilon,\infty}(\sigma,\tau)$.
    
    \begin{lmm}
      \label{lmm:ph2}
      All states $\sigma\not\in PH$ are transient for the Markov chain $P_{\varepsilon,\infty}(\sigma,\tau)$.
    \end{lmm}
    \begin{proof}
      Key to the proof is the following observation: for
      all initial configurations, there exists a finite probability
      $p>\varepsilon^{2L}$ for the system to be after $2L$ steps in the
      configuration $\sigma_{\text{queue}}$ such that $\sigma_i=0$ for
      $i=1,2,...,L$ and $\sigma_i=1$ for $i=L+1,L+2,...,2L$. Indeed, if
      the blockage stops the passage of particles for $2L$ times in a row,
      then the system will surely reach the configuration
      $\sigma_{\text{queue}}$, which is a state obviously ph-symmetric. 
      Hence, starting from any initial state
      $\sigma\not\in PH$, there is a finite probability that the system arrives in a
      symmetric state after $2L$ steps, and in virtue of Lemma~\ref{lmm:ph1} 
      it will never visit again $\sigma$.
    \end{proof}
    
    The next Lemma is motivated by the following easy yet important
    \begin{rmk}
      \label{rmk:freetomove}
      Let us imagine that after some time the system reaches a
      ph-symmetric configuration such that
      all particles in $\{1,2,...,L\}$ are free to move, then the
      particles will be free to move in that half of the circle at
      any subsequent time.  This happens because all the particles in
      $\{1,2,...,L-1\}$ can never reach the preceding particle (they all
      move at each step with probability $1$) and if $\sigma_L=1$ then
      $\sigma_{-L}=\sigma_{L+1}=0$, so that the presence of a particle at
      site $L$ cannot form a queue behind itself.
    \end{rmk}
    
    \begin{dfn}
      We will call $\Omega_\infty$ the set of configurations $\sigma\in PH$ such
      that all the particles in $\{1,2,...,L\}$ are free to move.
    \end{dfn}
    
    \begin{lmm}\label{lmm:omegainf}
      For the Markov chain $P_{\varepsilon,\infty}(\sigma,\tau)$ all the
      states $\sigma\not\in \Omega_\infty$ are transient.
    \end{lmm}
    
    \begin{proof}
      The proof is analogous to that of Lemma~\ref{lmm:ph2}.
      For any starting configuration $\sigma \not\in \Omega_\infty$, the system has a finite
      probability to reach the configuration $\sigma_{\text{queue}}$ in
      $2L$ steps, and $\sigma_{\text{queue}}\in \Omega_\infty$. The thesis
      then follows from Remark~\ref{rmk:freetomove}.
    \end{proof}
    
    A direct consequence of Lemma~\ref{lmm:omegainf} is that
    the stationary probability $\infPi$ of the Markov
    chain $\infP(\sigma,\tau)$
    is supported on $\Omega_\infty$, where the
    latter is manifestly ergodic.
    
    Let $\infJ = \lim_{L\to\infty}\infPi(\sigma_i=1,
    \sigma_{i+1}=0)$ be the current of parallel \ac{TASEP}
    in the regime of $\myw \to \infty$ under the action of a blockage $\varepsilon$, and let
    $r(\sigma)$ be the number of particles that lie in first half of the
    circle, i.e.\ sites $\{1,2,...,L\}$; those particles are
    free to move by Lemma~\ref{lmm:omegainf}. 
    It is quite obvious that the number of particles free to move
    in sites $\{L+1,L+2,...,2L\}$ is again $r(\sigma)$ by the ph-symmetry. 
    Then, analogously to what we have done in~\eqref{eq:Jpar}, we can
    compute $J_{\varepsilon,\infty}$ as 
    \begin{equation}\label{defbloc}
      J_{\varepsilon,\infty}=\lim_{L\to\infty}\frac{R}{L}\,,
    \end{equation}
    where $R = \mathbb{E}_{\infPi}(r)$ is the expected value of
    $r(\sigma)$ with respect to the stationary measure $\infPi$.
    
    We are now ready to prove the fundamental result of this section.
    \begin{prp}
      \label{prp:5}
      The current $\infJ$ of the Markov chain
      $\infP(\sigma,\tau)$ is
      \begin{equation}\label{bloc}
        \infJ = \frac{1-\varepsilon}{2-\varepsilon}\,.
      \end{equation}
    \end{prp}
    
    \noindent
    Key to the proof of Proposition~\ref{prp:5} is the following
    \begin{rmk}
      \label{rmk:bello}
      Starting from a state $\sigma \in \Omega_\infty$,
      the first $L$ sites loose any memory of the initial configuration
      after $L$ steps of the dynamics because the particles are free to
      move in the first half of the ring. Because of the ph-symmetry, $L$
      steps are sufficient also for the second half of the ring to loose
      any memory of the initial configuration.
      In other words,
      let $\sigma_t$ be the configuration of the chain after $t$ steps and
      let $\zeta_\infty$ be the \emph{hitting time} of the set
      $\Omega_\infty$, i.e.\ $\zeta_\infty = \min\{ t\geq0 \text{ such that } \sigma_t \in
      \Omega_\infty\}$. Then, $\zeta_\infty + L$ is a \emph{strong stationary
        time} fo rthe chain.
    \end{rmk}
    
    \begin{proof}[Proof of Proposition~\ref{prp:5}]
      The first part of the proof is the computation of the stationary
      measure of the chain.
      We can imagine that, at each step in which
      $\sigma_{2L}=1$, the blockage is driven by a binary random variable,
      which can be \emph{green}, giving at the next step $\tau_{2L}=0$, or
      \emph{red}, giving at the next step $\tau_{2L}=1$.
      Due to the
      symmetry, we can say that the probability of each state can be
      written in terms of a sequence of green and red lights. In
      particular, when the particle has passed the blockage, and therefore
      $\sigma_{2L}=0, \sigma_1=1$, we know that in the next step we will
      have for sure $\tau_{1}=0, \tau_2=1$. By symmetry, this means that
      we have now a particle in the site $2L$. Hence the next generation
      of a particle in the set $\{1,2,...,L\}$ is due only to the values
      of the red light.
      
      \noindent Let us first consider the set of states $\sigma\in\Omega_\infty$ such that $\sigma_1=0$.
      Then, if site $i$ is occupied then site $i-1$ is a hole with probability $1$.
      In other words, each particle \emph{occupies} two sites and so the
      probability\footnote{Actually, this is the stationary probability due to Remark~\ref{rmk:bello}.}
      to have a configuration $\sigma$ with $r$ particles lying in $\{1,2,...,L\}$ is
      \begin{equation}
        \label{probbloc}
        \pi_{\varepsilon,\infty}(\sigma)=(1-\varepsilon)^r\varepsilon^{L-2r}.
      \end{equation}
      Conversely, if $\sigma_1=1$ then the exponents appearing in~\eqref{probbloc}
      may increase or decrease by a unit at most. This correction is negligible
      in the thermodynamic limit, thus we will use~\eqref{probbloc} to compute the
      current according to~\eqref{defbloc}. For large $L$,
      \begin{equation}\label{R1}
        R \approx \frac{\sum_{r=1}^{L/2}r{L-r\choose r}(1-\varepsilon)^r\varepsilon^{L-2r}}
        {\sum_{r=1}^{L/2} {L-r\choose r}(1-\varepsilon)^r\varepsilon^{L-2r}},
      \end{equation}
      and we can again evaluate the expression simply by using the approximation
      in~\eqref{9.1} and the saddle-point method. 
      Let us call $x=\frac{r}{L}$ and recast~\eqref{R1} as
      \begin{equation}\label{R}
        R \approx
        L \frac{\int_0^{L/2 } dx\, x \, e^{L((1-2x)\ln\varepsilon+x\ln(1-\varepsilon)-x\ln\frac{x}{1-x}-(1-2x)\ln\frac{1-2x}{1-x})}} 
        {\int_0^{L/2} dx \, e^{L((1-2x)\ln\varepsilon+x\ln(1-\varepsilon)-x\ln\frac{x}{1-x}-(1-2x)\ln\frac{1-2x}{1-x})}}\,.
      \end{equation}
      In the limit of $L\to\infty$,
      \[\infJ = \lim_{L\to\infty}\frac{R}{L} = \bar{x}\,,\] 
      where $\bar{x} = \frac{1-\varepsilon}{2-\varepsilon}$ is the value of $x$ that maximises
      \[f(x)=(1-2x)\ln\varepsilon+x\ln(1-\varepsilon)-x\ln\frac{x}{1-x}-(1-2x)\ln\frac{1-2x}{1-x}\,. \]
    \end{proof}
    
    \begin{rmk}
      Despite its simplicity, this computation proves, in this completely
      parallel context, that a very small perturbation of the transition
      probabilities in a single site extends its effect over all the
      volume, without any fading. Indeed, the uniform density of the
      particles in the set $\{1,2,...,L\}$ is
      $\frac{1-\varepsilon}{2-\varepsilon}$ while in the set
      $\{L+1,L+2,...,2L\}$ the uniform density is
      $\frac{1}{2-\varepsilon}$.
    \end{rmk}

    \section{Numerical results}\label{num}
    
    In this final section we present a series of numerical evidences 
    obtained in a half-filled parallel \ac{TASEP}
    with blockage probability $\varepsilon$ acting on the site $2L$ 
    of a ring lattice of $1000$ sites ($L=500$).
    In particular, we numerically evaluate the current
    the current, $J(p,\varepsilon)$, and the particle density around site $x$, $\rho(x,p,\varepsilon)$,
    as functions of both the probability
    $p=\nicefrac{\myw}{(1+\myw)}$
    of an engine moving to the next site and the blockage intensity
    $\varepsilon$.
    The current is computed according to~\eqref{eq:Jpar}, while density
    (see Section~\ref{sec:density} below) is
    computed at points $x\in\{1,2,...,\nicefrac{2L}{10} \}$ as follows:
    \[
    \rho(x,p,\varepsilon)=\frac{1}{10}\left[ \sum_{i=10x}^{10x+9}\sigma_i\right].
    \]

    The problem of rigorously evaluating the mixing time of parallel \ac{TASEP} is not studied
    in this paper. However, the argument leading to the evaluation of the stationary measure
    in the case $p=1$ shows that the system reaches the stationary distribution
    after a time proportional to $L$.
    
    In the general case $p<1$ we run the dynamics for a time 
    \[
    T =2\ \frac{L}{p}~log(L)\,.
    \]
    We are not aware of an estimate of the mixing time for the standard \ac{TASEP}.
    In the case of the symmetric exclusion process on the circle Morris proved that 
    $T=L^2\log L$ (see~\cite{morris2006mixing}),
    and this corresponds in our case to the choice $p=\nicefrac{1}{2L}$.
    
    Our numerical experiments are particularly focused on two facts: 
    \begin{enumerate}
    \item We know from~\cite{costin2012blockage}
      that in serial \ac{TASEP} the current remains very close to the 
      limit value without blockage. 
      We also know that for the explicitly solvable blockage system
      discussed in Section~\ref{block}, i.e.\ $p=1$ or, equivalently,
      $\myw\to\infty$, the current has a decrease that is
      proportional to $\varepsilon$ near the value $\varepsilon=0$.
      We want to check numerically if the supposed non-analyticity of $J$
      around $\varepsilon=0$
      is a particular feature of the single spin-flip dynamic, or if it survives to the parallel case.
      The following figures clearly show that the second scenario seems the
      one to be true, at least from a numerical point of view.
    \item We want to see if the density is an increasing function of the
      site $x$ when the current decreases,
      or if the presence of the blockage implies simply a \emph{queueing}
      of particles close to it; the solvable model with $p=1$, i.e.\ $\myw\to\infty$, exhibits the
      latter behaviour.
    \end{enumerate}
    
    \subsection{Current}
    \label{sec:current}
    
    
    The measure of the current $J$ is simply made by counting the average number of the particles
    free to move during the evolution of the system and weighing such value with the total volume 
    of the system $2L$.
    
    
    \begin{figure}[tbp]
      \centering
      \includegraphics[width=\textwidth]{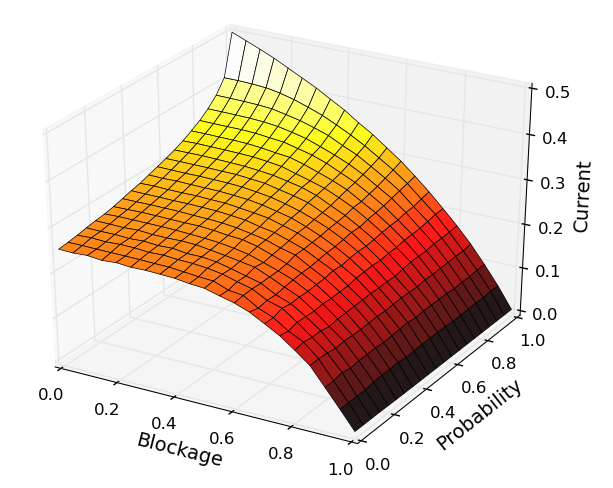}
      \caption{Current in\ parallel \ac{TASEP} as function of the probability
        $p=\nicefrac{\myw}{(1+\myw)}$ and the blockage intensity $\varepsilon$.}
      \label{fig:Jpe}
    \end{figure}
    
    \begin{figure}[tbp]
      \centering
      \includegraphics[width=\textwidth]{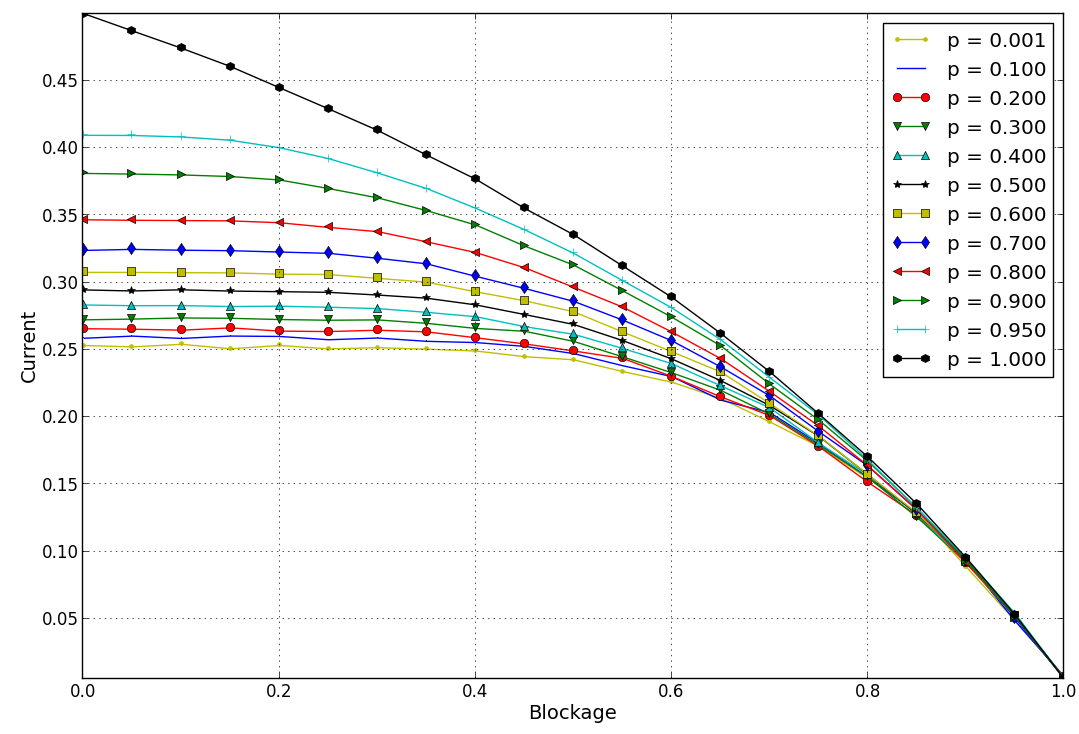}
      \caption{Profile curves of the current in the parallel \ac{TASEP}
        for different values of the probability $p = \nicefrac{\myw}{(1+\myw)}$.}
      \label{fig:Jslice}
    \end{figure}
    
    \begin{figure}[tbp]
      \centering
      \includegraphics[width=\textwidth]{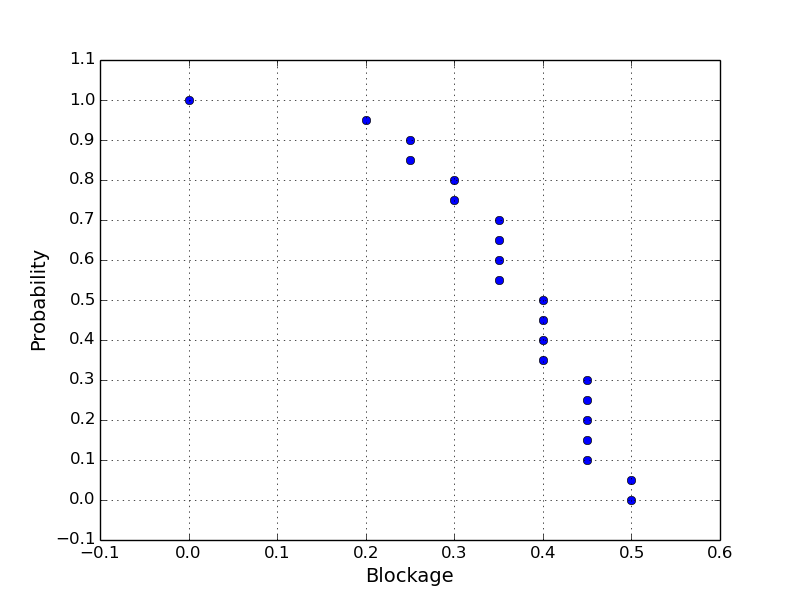}
      \caption{Threshold values of $\varepsilon$
        for a $1\%$ deviation from the no-blockage current.}
      \label{fig:Jtolerance}
    \end{figure}
    
    Figure~\ref{fig:Jpe} shows the surface obtained interpolating 441 measures of $J$ with every
    combination of the parameters $\frac{1}{2L} \leq p \leq 1$ and
    $0\leq \varepsilon\leq 1$, both with increments of $\nicefrac{25}{L} = 0.05$. 
    The figure clearly shows an excellent fit with the currents computed in Lemma~\ref{lmm:parJ}
    and in Proposition~\ref{prp:5}.
    
    Figure~\ref{fig:Jslice} presents the behaviour of the current for
    $\nicefrac{1}{2L} \leq p \leq 1$ with increments of $\nicefrac{25}{L} = 0.05$,
    plotting the side projection of the 3D graph
    from Figure~\ref{fig:Jpe}. 
    It clearly appears that except in the case $p = 1$, i.e.\ $\myw\to\infty$, where the current decrease with a
    finite slope for all $\varepsilon>0$, the decrease of $J$ starts only
    after a certain value of the blockage. 
    In this respect, the conjectured non-analytical behaviour of the serial
    \ac{TASEP}, to which the 
    parallel \ac{TASEP} corresponds in the regime $p = O(\nicefrac{1}{L})$,
    seems to be conserved for all the probabilities except $p = 1$.
    
    Figure \ref{fig:Jtolerance} plots the threshold values of $\varepsilon$
    for which the current deviates more than the $1\%$ from its initial value in the absence of blockage. 
    This gives an indication of the shape of the region in which the current remains nearly
    constant from a numerical point of view, having however $\varepsilon>0$.
    The description of this behaviour could be investigated after we had achieved a better
    understanding of the stationary measure in presence of blockage (see open questions below).
    
    \subsection{Density}
    \label{sec:density}
    
    Simulations can be also used to obtain insights about 
    the typical particle distribution over the ring.
    In order to obtain this characterisation, 
    we have coarse-grained the particles on segments of length 10, in order to
    obtain the density $\rho$ defined above.
    The density diagrams below describe the density of segments of the configuration, assigning a darker colour to the more dense segment. 
    
    \begin{figure}[tbp]
      \centering
      \includegraphics[width=\textwidth]{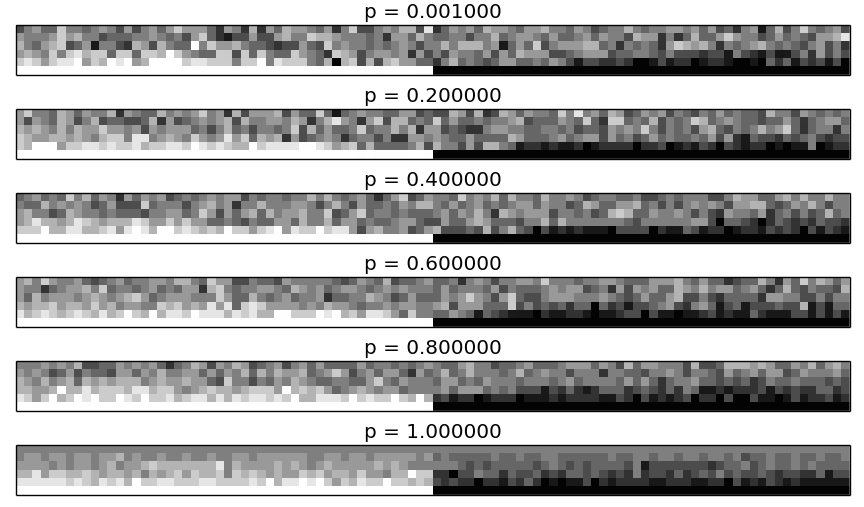}
      \caption{Point estimate of the density $\rho$ in a
        parallel \ac{TASEP} after $T = 2 \frac{L}{p}\,\log L$ steps. 
        The parameter $p$ ranges in $\{0.001, 0.2, 0.4, 0.6,
        0.8, 1.0\}$, while $\varepsilon$ ranges in $\{0, 0.2, 0.4, 0.6, 0.8, 1\}$.}
      \label{fig:DEN}
    \end{figure}
    
    \begin{figure}[tbp]
      \centering
      \includegraphics[width=\textwidth]{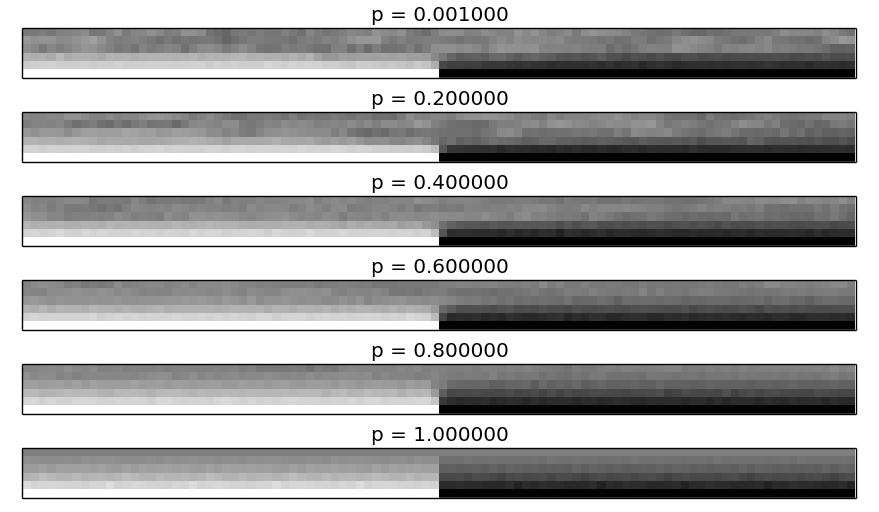}
      \caption{Mean density computed from 100
        replicas of the dynamics The density is computed from the
        states visited by the dynamics between time $T$ and $T+100$,
        where $T = 2 \frac{L}{p}\,\log L$. The parameter 
        $p$ ranges in $\{0.001, 0.2, 0.4, 0.6,
        0.8, 1.0\}$, while $\varepsilon$ ranges in $\{0, 0.2, 0.4, 0.6, 0.8, 1\}$.}
      \label{fig:EVO}
    \end{figure}
    
    
    The diagrams in Figure~\ref{fig:DEN} -- one for each value of $p$ --
    are instantaneous plots of the configuration $\sigma$ at the time 
    $T = 2 \frac{L}{p}\,\log L$. Each diagram is composed of 6 tiny rows
    showing the particle density for
    $\varepsilon \in \{0, 0.2, 0.4, 0.6, 0.8, 1\}$.
    As expected, the last row of every diagram is split in half white-empty
    and half black-full dots due the total congestion of the blockage at $\varepsilon=1$.
    
    Figure~\ref{fig:EVO} is analogous to Figure~\ref{fig:DEN} but it plots the average behaviour
    computed from $100$ density diagrams.
    Exactly like the previous graph, the first row corresponds to the complete absence of blockage,
    and its uniform colour fairly reflects the rotational invariance of the
    system.
    
    These evidences seems to indicate the possibility that, although the
    current decreases for all $\epsilon$ (see~\cite{basu2014last}),
    the system seems to exhibit for $p<1$ a nearly-constant density up to a certain
    value of $\epsilon$, while after this value the density appears to be
    smaller in the first $L$ sites than in the second $L$.
    Such a difference in the density in the two halves of the ring
    is precisely what has emerged in Section~\ref{block} for the rule-184 cellular
    automaton ($p=1$) with an  arbitrary blockage-intensity $\epsilon>0$,
    where the ph-symmetry plays a key role.
    Again, a complete understanding of this phenomenon 
    should be based on the knowledge of the stationary measure in 
    presence of a blockage. 
    
    The numerical evidences above show that there are many open interesting
    questions about our model, namely,
    \begin{itemize}
    \item prove that, in absence of
      blockage, the mixing time $\tau$ of the process is of the order of
      $\frac{L}{p}\,\log(L)$;
    \item show that except in the case
      $p=1$, i.e.\ $\myw\to\infty$, the probability to have a particle in a site is increasing
      along the circle starting from the blockage point;
    \item investigate, in the general case $p<1$, i.e.\ $\myw$ finite, the stationary measure
      of the system in presence of blockage. The last point may be
      possibly tackled via some perturbative approach
      with respect to the two cases that are completely known ($\varepsilon=0$ and $p=1$).
    \end{itemize}

    
    \addtocontents{toc}{\protect\vspace{\beforebibskip}}
    \addcontentsline{toc}{section}{\refname}    
    \bibliographystyle{siam}
    \bibliography{SLM}
\end{document}